# Spin-orbit concept of open-shell systems

Elena F. Sheka

Theoretical Physics and Mechanics Department
Peoples' Friendship University of Russia
117198 Moscow, Russia

**Abstract:** The paper presents a first attempt to look at the peculiarities of open-shell systems from the viewpoint of spin-orbit coupling. Contrary to common opinion of a negligible role of the issue for light element molecules, presented in the paper evidences its governing role in such systems peculiarities on example of $sp^2$ nanocarbons. A particular attention is given to the justification of the validity of UHF approach to exhibit the SOC peculiarities of open-shell systems.

1. **Introduction**

The nineteenth century was marked by the discovery of a highly peculiar molecule named by M. Faraday in 1825 as benzene which has become one of the pillars of the modern organic chemistry. The twentieth century was enriched in 1926 by the Hückel explanation of the benzene molecule peculiarity which has formed the grounds of the $\pi$-electron theory of aromaticity that was the ground of the modern quantum chemistry in general. The twenty first century has faced a conflict between compositions of condensed benzenoid units and the aromaticity theory. This conflict makes to abandon the benzene-aromaticity view on the compounds formed by the condensed benzenoid rings and stimulates finding conceptually new approaches for their description. The current paper is an attempt to answer this demand. The author suggests spin-orbit coupling to be laid in the foundation of the conflict resolution. The concept has arisen on the basis of a set of extended quantum chemical computational experiments performed by the author as well as by other scholars, results of which have found a convincing empirical support. In the way of the concept formation, one had to answer the question how nonrelativistic formalism of tools used in the course of the above computational experiments is able to exhibit pure relativistic features. The solution of this and other problems met in the way is described in the current paper. The paper consists of the following parts. Section 2 is devoted to the general characterization of open-shell molecules as the main objects of the concept. Section 3 contains empirical confirmation of the UHF peculiarities of open-shell molecules. Common nature of the $sp^2$ nanocarbons peculiarities is outlined in Section 4. Spin orbit coupling in quantum chemical calculations is considered in Section 5. Section 6 is devoted to the adequacy of UHF results to the SOC expectations. The evaluation of the SOC parameter of $sp^2$ nanocarbons on the basis of UHF results is presented in Section 7. Conclusion summarizes the paper essentials.

## 2. General characteristics of open-shell molecules

The term 'open-shell molecule' covers a large set of species differing quite considerably. It was firstly attributed to radicals while generalizing the Roothaan iterative method of determining LCAO molecular orbitals when the number of α electrons is not equal to the number of β electrons ($N_\alpha \neq N_\beta$) [1] thus introducing the unrestricted Hartree-Fock approach into quantum chemistry. For a long time the term was associated with radicals or molecules with odd number of electrons. Fifty years after the first introduction, the term was applied to open-shell singlet diradicals based on olygocenes ($N_\alpha = N_\beta$).[2] At the same time the application of the unrestricted Hartree-Fock (UHF) formalism to singlet-ground-state fullerenes revealed their polyradical character [3-5] thus attributing the latter to the family of singlet open-shell molecules. Singlet molecules form a particular class, inside of which a special role belongs to $sp^2$ nanocarbons, such as polycyclic aromatic hydrocarbons [2, 6-17] (referred to as either linear acenes or polyacenes and olygoacenes), fullerenes [3-5, 18-25], carbon nanotubes, [26-29] graphenes, [30-42] graphdyines, [43] and so forth. Cited to as singlet biradicals, polyradicals or open-shell molecules in general, the molecules present the first case in molecular physics when the species were distinguished not at experimental but computational level. This action is intimately connected with spin-unrestricted formalism, mostly UHF one.

If a lower energy solution can be found by releasing the constraint that electrons of opposing spin occupy the same spatial orbital, the restricted Hartree-Fock (RHF) wavefunction (wf) is said to have triplet instability. In the spin unrestricted Hartree-Fock (UHF), different spatial orbitals are used for α and β spin. A further destiny of the UHF solution depends on whether the UHF wf is an eigenfunction of the total spin-squared operator $\hat{S}^2$ or not. Consequently, applying UHF approach to open-shell molecules, two classes of the latter can be distinguished: UHF solutions of the first class molecules are identical to those of either restricted open Hartree-Fock (ROHF) ($N_\alpha \neq N_\beta$) or RHF ($N_\alpha = N_\beta$) (see a number of numerous examples in [44]). The solution is spin pure and the UHF wf satisfies the total spin-squared operator $\hat{S}^2$ as well. In contrast, the UHF solutions for the second class molecules drastically differ from both ROHF and RHF ones. The difference consists in lowering energy and a remarkable spin contamination that is a result of the UHF wf not to be more the eigenfunction of the operator $\hat{S}^2$. The two classes of UHF solutions follow from the inner logic of the UHF formalism and are dependent on molecular object under consideration. [45]

Not paying attention to the latter, the spin contamination of the UHF solutions, which casts doubts on the purity of spin multiplicity of the molecule ground state, quite often is attributed to the method disadvantage that leads to erroneous results (see [46] and references therein) due to which singlet diradicals are classified as problematic species. [47] At the same time, the UHF spin contamination is usually considered as pointing to enhanced electron correlation that requires configuration interaction (CI) schemes for its description. Argued, that if UHF approach as the first stage of the CI ones is improved towards a complete CI theory, one would expect removing the spin contamination.

However, two extended computational experiments performed with interval of 15 years fully discard this expectation. The first was carried out back in 2000 on 80 molecules studying the HF solution instability. [48] The molecule set covered a large variety of species (see Chart 1) including valence saturated and unsaturated compounds, fully carbonaceous and containing heteroatoms. The instability feature was studied in relation with the electronic correlation, the vicinity of the triplet and singlet excited states, the electronic delocalization linked with resonance, the nature of eventual heteroatoms, and the size of the systems. It was shown, that for most conjugated systems, the RHF wf of the singlet fundamental state presents so-called triplet instability that differs by value. [49] The largest effect was observed for aromatic hydrocarbons.

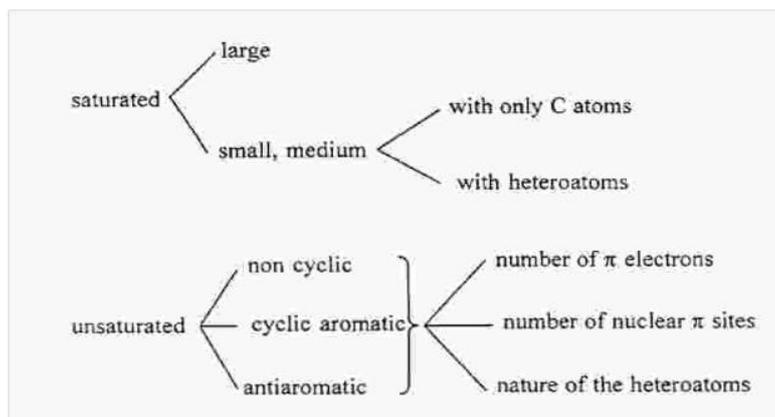

Chart 1

The second experiment has been performed just recently for 14 polyaromatic hydrocarbons by using a number of different CI approaches such as UHF, UMP2, QCISD(T), and UDFT. [16] Results, concerning spin contamination $\langle \Delta S^2 \rangle$ of the molecules, are well consistent for the first three techniques while the latter in the case of UDFT was practically null. Actually, the data are dependent on the approach in use. However, the difference within either HF- or DFT-based CI approaches occurred to be not as big as that between the HF and DFT approaches of the same level. The first consequence is due to the fact that triplet states are mainly responsible for the HF instability, thus allowing a considerable wf truncation. At the same time, the UHF formalism is fully adapted to the consideration of the triplet instability [50] due to which the UHF results deviate from the higher approaches of the CI theory no more than 20%. Once wf-based, HF consideration is preferential since the DFT one is much less adapted to the consideration of delicate peculiarities connected with the dynamic correlation of electron of different spins (see fundamental Kaplan's comments [51] and the latest comprehensive review [52]). In support of the said above, Fig. 1 portrays the data related to the total number of effectively unpaired electrons $N_D$ which can be considered as a qualitative measure of the spin contamination and which for the singlet state is [53]

$$N_D = 2\Delta \hat{S}^2. \tag{1}$$

Here, $\Delta \hat{S}^2$ is the deviation of squared spin from the exact value due to spin contamination. Presented data are related to a number of olygoacenes and were obtained by using both UHF semi-empirical codes [54] and density matrix renormalization group (DMRG) algorithms [6].

According to the UHF AM1 algorithm implemented in the CLUSTER-Z1 codes, used in the current study,

$$N_D = 2\left(\frac{N^\alpha + N^\beta}{2} - \sum_{i,j=1}^{NORBS} P_{ij}^\alpha * P_{ij}^\beta\right). \tag{2}$$

Here $N^\alpha$ and $N^\beta$ are the numbers of electrons with spin α and β, respectively, while $P_{ij}^\alpha$ and while $P_{ij}^\beta$ are the matrix elements of the relevant electron density matrices. The summation in Eq. 2 runs over all spinorbitals.

In appliance with the DMRG algorithm, [6] $N_D$ is determined as

$$N_D = \sum_i n_i (2 - n_i). \tag{3}$$

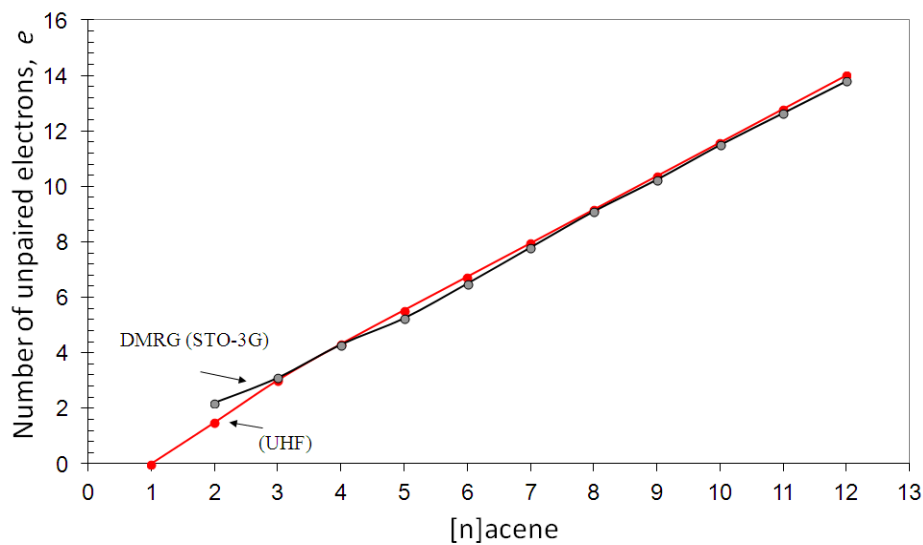

**Figure** 1. Total number of effectively unpaired electrons $N_D$ in polyacenes calculated by using DMRG (STO-3G))[6] and UHF (current paper) formalism. No scaling of the data.

Here, $n_i$ is the occupation number of the *i*th natural orbital that ranges from 0 to 2. The derivation of both Eq. 2 and Eq. 3 corresponds to the same basic concept on effectively unpaired electrons suggested in.[53, 55] As seen in Fig. 1, thus obtained $N_D$ values well coincide in both cases evidencing that the UHF capacity is quite high to be used as an computationally affordable approach that treats both static and dynamic correlations in a balanced manner. Similar picture is obtained in all the cases where UHF data can be compared with those obtained in the framework of higher-level CI approaches. Oppositely, in all known cases UDFT leads to underestimated data that are in conflict with empirical ones as it is in the case of the description of diradical character of the Cope rearrangement transition state.[56] When UDFT calculations gave $N_D$ = 0, CASSCF, MRCI, and UHF calculations gave 1.05 *e*, 1.55 *e*, and 1.45 *e*, respectively. Thus, experimentally recognized radical character of the transition state was well supported by the latter three techniques with quite a small deviation in numerical quantities while UDFT just rejected the radical character of the state. Is untrue as well that according to UDFT the radicalization of *n*-oligoacenes starts from heptacene[2] while the radical activity is actually observed even for naphthalene. Such examples can be numbered (see for example [57] and references therein).

### 3. Reality of UHF peculiarities of open-shell molecules

In spite of the fact that UHF triplet instability resulting in spin contamination is characteristic for particular open-shell molecules only, as said early, a considerable part of the scientific community still tends to opinion that this peculiarity is mainly attributed to drawbacks of the computational methods used. However, there are a lot of experimental evidences that the open-shell molecules peculiarities are real. Thus, addressing to Fig.1, it is necessary to remain the hampered availability of longer acenes, with pentacene being the largest well-characterized acene. In recent years substantial progress has resulted in the synthesis of n-acenes up to n=9 by matrix isolation techniques (see Ref. [10,39] and references therein). Nevertheless, these higher acenes are very reactive; for example, heptacene was found to be stable only for 4 h in a poly(methyl methacrylate) matrix. To overcome the stability problems, larger acenes were functionalized by adding protecting groups which inhibit the native high reactivity of the acenes.

In addition to the indirect $N_D$ manifestation discussed above, the last decade has provided convincing direct evidences of the $N_D$ existence. Figure 2b presents AFM atom-resolved images of two $sp^2$ molecules recorded in Zürich Research Laboratory of IBM Research.[58, 59] The first image portrays pentacene while the second is related to the smallest possible five-ringed olygoacene named olympicene in commemorating London's Olympic Games 2012. The two images are accompanied with the calculated distributions of effectively unpaired electrons over the molecules atoms [37] (so called $N_{DA}$ maps where $N_D = \sum_A N_{DA}$ [21]). According to the experimental set up, in the discussed experiment the AFM image brightness is the highest on the atoms of two edge pairs with the least force of the attraction the oxygen tip atom. Evidently, this should correspond to the least chemical activity of the atoms. In contrast, brightness on the calculated molecule portraits is the highest at the central atoms that are the most active and are characterized by the largest $N_{DA}$ due to which experimental image and calculated maps should be strongly color-inverse, which is really seen in Fig. 2a.

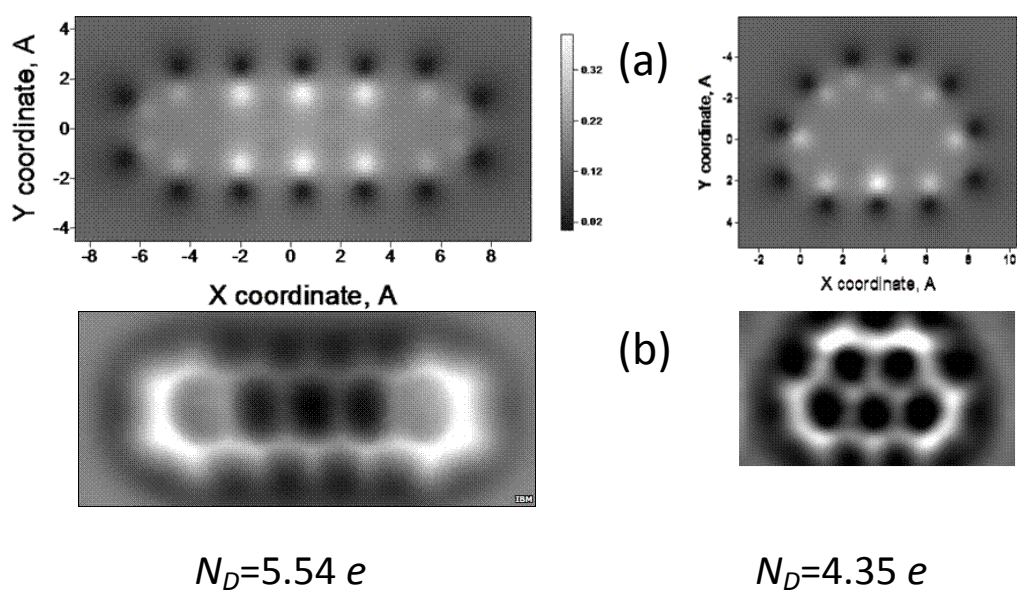

$N_D$=5.54 $e$          $N_D$=4.35 $e$

**Figure 2**. a. The $N_{DA}$ distribution (UHF calculation) over pentacene (left) and olympicene (right) molecules atoms. [37] b. AFM imaging of $sp^2$-open-shell molecules [58, 59] (by kind permission of L.Gross).

The other quite numerous observations of effectively unpaired electrons are related to graphene. The first case concerns graphene bubbles. Raised above the substrate and mechanically deformed areas of graphene in the form of bubbles are found on different substrates. They come in a variety of shapes, including those which allow strong modification of the electronic properties of graphene. [60, 61] Typically, the bubbles are seen as bright spots on dark background formed by not wrinkled graphene film. The bright spots of the experimental images exhibit places of graphene film with the largest electron density. Obviously, the density excess is caused by the film local deformation (curving) which is usually accompanied with increasing the number of effectively unpaired electrons. [62, 63] Since $N_{DA}$ values strongly depend on C-C distances formed by the atom [43] the strain-induced stretching of curved graphene bonds evidently causes the value enlarging that is revealed as enhanced brightness of the AFM images in the figure. The concept on strain-induced pseudomagnetic field suggested by the authors [60] and further considered in [64] follows just other way to present the bond stretching via the description of their elastic strain in terms of the effective electromagnetic field suggested in. [65]

Since $N_D$ is the measure of not only spin contamination but the extent of the relevant molecules radicalization, [18-21] locality of chemical reactions can indicate particular places with enhanced $N_{DA}$ values. Thus, a peculiar picture of decoration of graphene reactivity centers with Pd clusters [68] demonstrates possibility to obtain the spatial information about chemical reactivity across the Pd/C system. A particular role of wrinkle as nanosize gas-inlet for reactions under graphene is shown in. [69] The corrugated graphene has been recently observed as a covering over gold nanoparticles situated at a substrate. [66] Convexity enhanced brightness of graphene spots located over the gold particles is clearly seen.

The list of empirical evidences proving the existence of effectively unpaired electrons, responsible for a peculiar magnetism of *sp²* species that will be considered in Section 6, in particular, can be continued. However, already discussed is quite enough to accept this issue of physical and chemical reality. Therefore, computational findings and physico-chemical reality do not contradict and *sp²* open-shell molecules (*sp²*-OSMs below) do reveal particular properties. Summarizing, the latter can be formulated as following:

1. Singlet ground state of *sp²*-OSMs is really spin-contaminated, which means that its singlet spin multiplicity is not exact;
2. Spin contamination is accompanied by the appearance of effectively unpaired electrons that are an extent measure of both spin-contamination and enhanced chemical activity and/or radicalization of the molecules;
3. Spin contamination is remarkably strain dependent indicating the crucial role of C-C spacings.

Common for all the *sp²*-OSMs and not dependent on chemical content, shape, and size of the latter, the properties evidently have a common nature that has two faces, namely, empirical, connected with peculiar structures of the molecules, and theoretical, implemented in the UHF formalism.

## 4. Common nature of the *sp²*-OSMs peculiarities

As shown by the author studies of recent years, [28, 29, 31, 32, 35, 36, 40, 43] the C-C bonds length and its variety is the cradle of the *sp²*-OSMs peculiarities. When the bond lengths exceed a critical value $R_{crit}$ = 1.395Å, RHF-character of the UHF solution is transferred to the spin-contaminated UHF one. Lengths of a considerable part of C-C bonds of *sp²*-OSMs, such as polyacenes, fullerenes, CNTs, and graphene, are above the critical value. This characteristic property of the UHF solutions, concerning the RHF/UHF transformation depending on the length of the covalent bond under consideration, is well known. [45] The results of computational experiments [16, 48] over large sets of open-shell molecules, analyzed from this viewpoint, perfectly support this conclusion as well.

Apparently, bond-length concept, so well exhibited by the UHF formalism, could be considered as theoretical explanation of the UHF peculiarities of *sp²*-OSMs. However, the concept does not explain the origin of the main feature of the solutions – their spin contamination. Evidently, the reason should be sought among fundamentals that so far have been ignored at the level of the reference quantum molecular theory. The latter should be attributed to the quantum chemical description based on a nonrelativistic Hamiltonian, which describes the motion of electrons in central Coulomb field, and forms the ground of numerous restricted single-determinant (RSD) computational tools. As shown above, none of these tools is a proper approach in case of *sp²*-OSMs for which conceptually new approaches are necessary to go beyond the benchmark. Before speaking about the approaches, let us look what is ignored at the reference level.

It is convenient to represent expected progress in improving calculations from the RSD approach as a series of correction terms to the initial RSD energy $E_{RSD}$

$$E_{best\ estimate} = E_{RSD} + \Delta E_{corr} + \Delta E_{ScR} + \Delta E_{SO} + \Delta E_{BOC} + \Delta E_{ZPE} + \cdots \quad (4)$$

The main corrections cover correlation energy $\Delta E_{corr}$, scalar-relativistic $\Delta E_{ScR}$ and vectorial-relativistic spin-orbit $\Delta E_{SO}$ contributions, non-Born-Oppenheimer $\Delta E_{BOC}$ correction, zero-point vibrational effect $\Delta E_{ZPE}$, and other much less significant corrections intimately connected with nuclear motion. Taking into account each of these terms may considerably change the RSD results, which is confirmed by a profound development of quantum chemistry over the last century. However, only spin-orbit coupling (SOC) can mix states of different spins. But a straight inclusion of SOC into consideration means passing to a valley under supervision of relativistic quantum chemistry. What then one has to do with nonrelativistic UHF and other approaches, the use of which gives a clear vision of spin mixing and provides results in excellent agreement with experimental data? Looking for answering this question with respect to UHF, two other questions arise: 1) Do really the UHF features look like those caused by SOC? As will be shown below, answer to the question is positive. Consequently, the second question stems: How SOC peculiarities can be represented by nonrelativistic computational tool?

It should be said that when the author's SO concept of the UHF approach was formulated [70] and the current paper was nearing completion, it was still a question of disclosing a direct connection between the UHF formalism of molecular theory with the relativistic one due to which only directions where the connection should be sought could be indicated. One of such directions concerns the analogy of the perturbation theory (PT) treatment of both UHF and SOC approaches, which will be considered in the next section. As for general molecular theory, fortunately, just recently two important papers have been found. The first was written by Deharend and Dive fifteen years ago [48] where an analogous view on the spin mixing of ground states of OSMs was expressed. However, the authors limited themselves by the idea only which did not obtain a further development. The second paper appeared in this year August. [71] The authors considered the spin-contamination problem in the course of a comparative study exploiting both UHF and Kramers-pairs-symmetry CGHF approaches of the relativistic molecular theory and for the first time showed a full analogy of the data obtained by the two formalisms. This fundamental study tightly connects nonrelativistic UHF formalism with relativistic molecular theory and allows successfully proceeding with the SOC origin of the UHF peculiarities of OSMs.

## 5. SOC and quantum chemical calculations

### 5.1. General characteristics of spin-orbit coupling in molecules

The SOC in molecules consisting of light elements has been known quite long ago (for early works see [72] and references therein) as well as been a leitmotif of numerous theoretical investigations. Comprehensive reviews [73-78] and the second edition of well known textbook [79] are a must-have for everyone entering the field. With respect to experimental evidence, SOC in molecules shows itself via the following issues:
1. Splitting of the molecule ground state by removing spatial and/or spin degeneracy;
2. Lifting the ban of spin-forbidden transitions, both radiative and non-radiative;
3. Enhancing the molecule chemical activity caused by their radicalization;
4. Exhibiting a particular molecular 'para- and ferrodiamagnetism' (in terms of [80]).

The first issue is connected with the fact that SO couples total orbital moment of molecule *L* with its total spin *S* thus depriving them both of the quality of good quantum numbers. Such a role is transferred to the total momentum *J* with components *-(L+S); -(L+S)+1;… 0….(L+S)-1; L+S*. Consequently, the ground state is split into the relevant *J* components. Evidently, this causes lowering the ground state energy as a whole with respect to the RSD degenerate case. As for the splitting itself, for light-element molecules it does not exceed a few meV due to which its direct observation is possible if only the molecule rotational spectra are well resolved.

The governing role of the total momentum *J* is manifested in the fact that the same *J* component can correspond to different combinations of *L* and *S* thus providing a mixture of spin states at fixed energy. Such a mixture provides the appearance of issues 2-4, although in different way in each of them. It is necessary to say that in all the studied cases the mixture concerned singlet-triplet one. Thus, issue 2 is usually considered as a result of the mixture of the lowest triplet state either with singlet ground states (optical transitions causing the molecule phosphorescence) or with the singlet excited state to provide the radiationless relaxation of the latter, which is consistent with a particular role of triplet states in variable photophysics and photochemistry of molecules. Issue 3 is related to the mixture of singlet ground state with the lowest triplet one, which leads to its spin contamination. Peculiar distribution of spin density, caused by the above mixture, results in a variable response of molecules to the application of magnetic field, which lays the foundation of issue 4.

### 5.2. SOC, UHF and perturbation theory

Spin–orbit coupling arises naturally in Dirac theory, which is a fully relativistic one-particle one for spin 1/2 systems in electric field, [81] and is described by the SOC contribution in the total Hamiltonian in the approximate form when the SOC smallness is fulfilled

$$\hat{H}^{SO} = -i\frac{e\hbar^2}{8m^2c^2}\,\boldsymbol{\sigma}\cdot\boldsymbol{\nabla}\times\mathbf{E} - \frac{e\hbar}{4m^2c^2}\,\boldsymbol{\sigma}\cdot\mathbf{E}\times\mathbf{p}. \qquad (5)$$

Here σ and E are Pauli spin matrix and electrostatic potential, respectively. The first and second terms in the right-hand part of the equation describe scalar and vectorial SOC contributions. The smallness of the SOC contribution allows considering $H^{SO}$ as an additive to the total Hamiltonian of a many-electron system in the form

$$H_{tot} = H_{RSD} + H^{SO}. \qquad (6)$$

Here $H_{RSD}$ is a spin free nonrelativistic RSD Hamiltonian mentioned earlier. The main computational approach concerns this very Hamiltonian, selecting it within either Hartree-Fock [73, 74, 77] or density functional theory [79] and equation-of-motion approach, [78, 79] and treating SOC either perturbationally or variationally. Additional problem consists in a proper choice of the most suitable effective presentation of Hamiltonian $H^{SO}$. Reviews [73, 75] provide much food for thought on this issue.

For the further discussion concerning UHF it is worthwhile to pay a particular attention to a perturbation formalism of the SOC computation. As suggested in, [82] SOC appears in the second order of the perturbation theory (PT) applied to $H_{RSD} = H^0$ Hamiltonian

$$(H^0 - E_I^0)\Psi_I^1 = H^{SO}\Psi_I^0. \tag{7}$$

$$\Psi_I = \Psi_I^0 + \sum_{J(\neq I)}^{L} \frac{\langle \Psi_J^0 | H^{SO} | \Psi_I^0 \rangle}{(E_J^0 - E_I^0)} \Psi_J^0$$

$$\equiv \Psi_I^0 + \Psi_I^1.$$

Here $E_I^0$ and $\Psi_I^0$ are eigen value and eigen wf of spin free Hamiltonian $H^0$; $I$ labels different electronic states. The suggested formalism is widely used for determining governing SOC parameters (see [73] and references therein).

It is the very time to remain that UHF formalism straightly follows from perturbationally considered RHF one. [50, 83, 84] In this case Eq. 6 looks like

$$H_{UHF} = H_{RHF} - U. \tag{8}$$

Here $U$ is the spin-polarization operator determined by the difference of $H_{RHF}$ and $H_{UHF}$ operators. [84]

The study of Rosski and Karplus [83] was undertaken to obtain a proper formalism for describing spin effects in many-electron systems. They showed that when the UHF wf is presented as the first PT order to the RHF one (similarly to Eq. 7) lowering the RHF energy, square spin and spin density are strictly obtained in the second order of the perturbation theory in a perfect consent with the exact UHF solution.

Since the theory of RHF and UHF is well developed, it is possible to determine operator $U$ in terms of a familiar Fock operator. According to the fundamental paper of Pople and Nesbet, [1] the UHF wf is two-reference once distinguished for α and β electrons located at different spatial orbitals. Caused by the fact, two Fock operators $f^\alpha$ and $f^\beta$, related to electron (1) with either α or β spin, determine the UHF formalism:

$$f^\alpha(1) = h(1) + \sum_a^{N^\alpha}[J_a^\alpha(1) - K_a^\alpha(1)] + \sum_a^{N^\beta} J_a^\beta(1)$$

and (9)

$$f^\beta(1) = h(1) + \sum_a^{N^\beta}\left[J_a^\beta(1) - K_a^\beta(1)\right] + \sum_a^{N^\alpha} J_a^\alpha(1)$$

In the top equation $h(1)$ presents one-electron part of the operator while $J_a^\alpha(1)$ ($J_a^\beta(1)$) describes two-electron Coulomb interaction of a selected α electron with all other α (β) electrons and $K_a^\alpha(1)$ corresponds to exchange of the selected electron with all other α ones. Similar notations related to a selected β electron compose the bottom Eq. 9. The RHF solution is governed by the Fock operator in the form

$$f(1) = h(1) + \sum_a^{N/2} J_a(1) - K_a(1).$$

Accordingly, the difference operators $\Delta f^\alpha(1) = f^\alpha(1) - f(1)$ and $\Delta f^\beta(1) = f^\beta(1) - f(1)$ in the case when $N^\alpha = N^\beta = N/2$ can be expressed as

$$\Delta f^\alpha = -\sum_a^{N^\alpha} K_a^\alpha(1) + \sum_a^{N^\beta} J_a^\beta(1)$$

and (10)

$$\Delta f^\beta = -\sum_a^{N^\beta} K_a^\beta(1) + \sum_a^{N^\alpha} J_a^\alpha(1)$$

Since exchange members are usually small, the difference between RHF and UHF formalism is governed by nondiagonal elements of Coulomb interaction.

Therefore, we are facing the situation when two different operators $H^{SO}$ and $U$, both aimed at the description of intimate spin effects, are of conceptually different origin, namely, of relativistic and nonrelativistic, respectively. At the same time, as will be shown in Section 7.2, the treatment of operator $H^{SO}$ in terms of the UHF resultant data provides the SOC parameters well consistent with available experimental data.

### 5.3. SOC, UHF and relativistic molecular theory

The one-electron Dirac equation has the form

$$\hat{h}_D \Psi = E\Psi \tag{11}$$

where

$$\hat{h}_D = c\hat{\alpha} \cdot \hat{p} + \beta c^2 = \begin{pmatrix} c^2 I & c\hat{\sigma} \cdot \hat{p} \\ c\hat{\sigma} \cdot \hat{p} & -c^2 I \end{pmatrix} \tag{12}$$

and

$$\Psi = \begin{pmatrix} \Psi^L \\ \Psi^S \end{pmatrix} = \begin{pmatrix} \Psi^{L\alpha} \\ \Psi^{L\beta} \\ \Psi^{S\alpha} \\ \Psi^{S\beta} \end{pmatrix}. \tag{13}$$

Here, $\hat{\sigma}$ and $I$ are Pauli spin matrices and 2x2 identity matrix, $\hat{p}$ is pulse operator; $\Psi^L$ and $\Psi^S$ are two-component spinors in the Dirac four-spinor presentation containing large (L) and small (S) component functions, respectively. Hamiltonian of many-electron system, based on the above four-component one-electron one, is very complicated and presents high difficulties for its computing. Accordingly, a number of approximations have been suggested to make the problem solution feasible (see [74] and references therein).

To extend the Dirac equation for many–electron systems the proper relativistic electron–electron interaction is to be taken into account. The straightforward approach is to augment the one–electron Dirac operator with a two–electron interaction term. In the case of the most important Coulomb interaction, this term can be derived from quantum electrodynamics but cannot be written in a closed form. Therefore, in practice one prefers truncation to the so called Breit operator,[74]

$$g_{ij}^{Breit} = -\frac{1}{2r_{ij}}\left[\hat{\alpha}_i \cdot \hat{\alpha}_j + \frac{(\hat{\alpha}_i \cdot r_{ij})(\hat{\alpha}_j \cdot r_{ij})}{r_{ij}^2}\right] \qquad (14)$$

which is added to the Coulomb interaction term, $\frac{1}{r_{ij}}$. The first term in brackets on the right–hand side of Eq. 14 is the Gaunt interaction term, often used as an approximation of the Breit operator since it includes the largest portion of the full Breit form. The Gaunt operator contains spin–spin, orbit–orbit and spin–orbit contributions, while the second term in Eq. 14 is the gauge interaction representing retardation effects induced by the finite speed of light.

Taking the Breit operator into account, the Dirac-Coulomb-Breit/Gaunt Hamiltonian can be obtained in the form

$$\hat{H}_{DCB/DCG} = \sum_i^n \hat{h}_D(i) + \frac{1}{2}\sum_{i\neq j}^n \hat{g}_{ij}^{Breit/Gaunt} \qquad (15)$$

The one-electron operator $\hat{h}_D(i)$ is identical with the operator

$$\begin{bmatrix} \hat{V} & c\hat{\sigma}\cdot\hat{p} \\ c\hat{\sigma}\cdot\hat{p} & \hat{V} - 2c^2 \end{bmatrix},$$

where the external potential V is provided by static atomic nuclei.

In comparison to the nonrelativistic many–electron RSD Hamiltonian, $\hat{H}_{DCB/DCG}$ is the one–electron operator that makes the difference which is sometimes accompanied with the $\hat{g}_{ij}$ electron–electron interaction term in its relativistically specified form. Implementing within two-electron integral generating codes, the $\hat{H}_{DCB/DCG}$ operator can be transformed into the Dirac-Fock operator that has the same structure as its nonrelativistic UHF analogue

$$\hat{f}_{DHF}(i)\phi(i) = \varepsilon_i \phi(i). \qquad (16)$$

Here, $\phi(i)$ is the four–component-spinor representation of the one–electron state.

As shown recently,[71] not only Fock-like Eq. 16 itself is analogous to the UHF one, but its solution can be presented in the UHF terms, such as spin contamination expressed via $\Delta S^2$ and spin density over molecule atoms. Related to $C_2H_5O$, $\Delta S^2$ constitutes 0.6024, exactly the same for UHF and general complex Hartree-Fock (GCHF) method. The GCHF spin contamination was provided with the Kramers pairs symmetry breaking. Besides spin contamination, the GCHF solution represents the spin density distribution over the molecule atoms, which is presented in Fig. 3. As seen in the figure, the distribution patterns are fully identical in the case of UHF and GCHF formalisms while considerably different in the ROHF and UDFT cases. [71] These results alongside with comments given at the end of Section 2 give good reason not to consider UDFT formalism as a proper technique for treating spin effects in open-shell molecules.

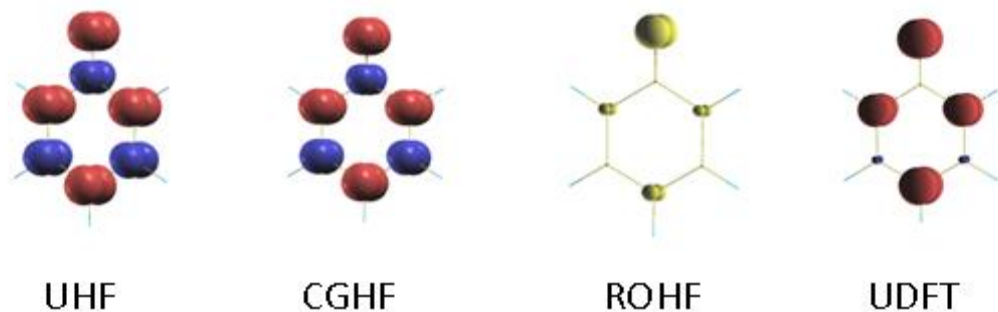

**Figure 3.** Spin density distribution over atoms of phenoxyl radical ($C_6H_5O$) calculated by using different formalisms.[71] (by kind permission of L. Bučinský).

## 6. Adequacy of the UHF results to the SOC expectations

The phenomenology of the UHF peculiarities of $sp^2$-OSMs comes to the following three issues:

**Issue 1**

$$\Delta E^{RU} \geq 0,$$

where

$$\Delta E^{RU} = E^R - E^U \tag{17}$$

presents a misalignment of energy of RHF ($E^R$) and UHF ($E^U$) solutions.

**Issue 2**

$$\Delta \hat{S}^2 \geq 0. \tag{18}$$

Here,

$$\Delta \hat{S}^2 = \hat{S}_U^2 - S(S+1)$$

is the misalignment of squared spin. $\hat{S}_U^2$ is the UHF squared spin while $S(S + 1)$ presents the exact RHF value of $\hat{S}^2$.

**Issue 3**

$$N_D \neq 0.$$

Here, $N_D$ is the total number of effectively unpaired electrons. The number is determined as

$$N_D = tr D(r|r') \neq 0 \quad \text{and} \quad N_D = \sum_A D_A. \tag{19}$$

$D(r|r')$ and $D_A$ present the total and atom-fractioned spin density caused by the spin asymmetry due to the location of electrons with different spins in different spaces.

In view of empirical SOC phenomenology outlined in Section 5.1, the first issue is a consequence of the SO splitting of the degenerate RHF states. Presented in Fig. 4 are selected sets of energies of HOMO and LUMO orbitals (25 in total) related to fullerene $C_{60}$, rectangle graphene molecule (5, 5) NGr with five benzenoid units along the armchair and zigzag edges, respectively, and a fragment of a bare (4, 4) single-wall CNT. The data are obtained by using RHF and UHF versions of the CLUSTER-Z1 codes. [54]

High degeneracy of the RHF solution of $C_{60}$ (Fig. 4a) is caused by both high spatial ($I_h$) and spin symmetry. According to the general SOC law, the orbitals are clearly split, which causes lowering the molecule spatial symmetry to $C_i$ while conserving the identity of spinorbitals related to α and β spins. The splitting value is quite remarkable resulting in the energy lowering of each HOMO orbital. Obviously, summed over all orbitals, the total lowering easily reach $\Delta E^{RU}$ = 14.82 kcal/mol.

Data presented in Fig. 4b are related to (5, 5) NGr molecule. The space symmetry of the molecule ($D_{2h}$) remains unchanged when going from RHF to UHF formalism. Degeneracy of the RHF orbitals and splitting of the UHF ones exhibits breaking spin symmetry that causes a remarkable distinguishing of orbitals related to α and β spins and is a direct consequence of the SOC. The energy lowering, summing over all orbitals, constitutes 307.5 kcal/mol.

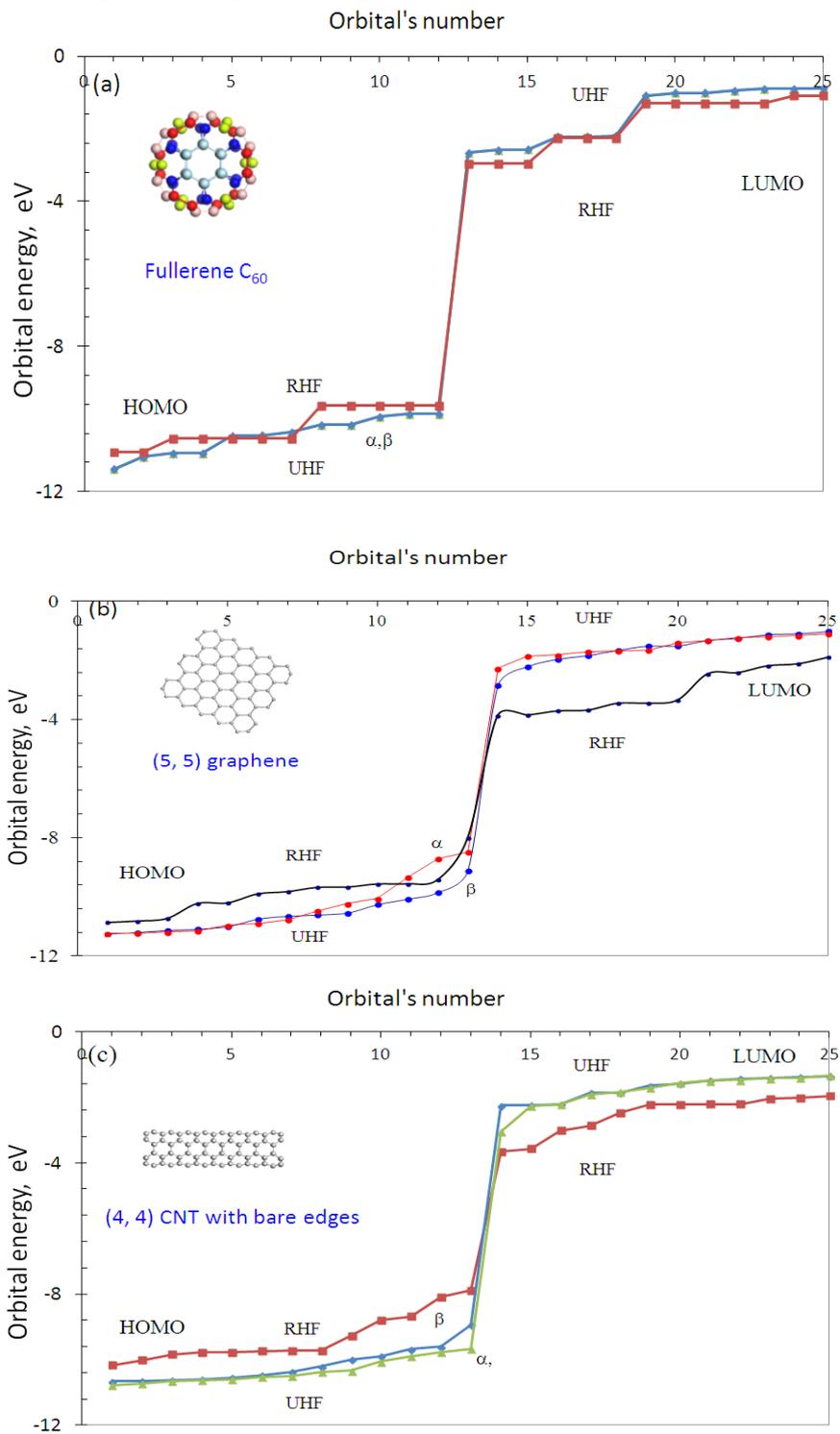

**Figure 4**. Energies of 25 spinorbital in the vicinity of HOMO-LUMO gap of fullerene $C_{60}$ (a), (5, 5) NGr molecule (b), and (4, 4) single-walled CNT with bare edges (c) (UHF, current paper).

Fully analogous picture is observed for the (4, 4) SWCNT fragment shown in Fig. 4c. Lowering the molecule space symmetry from $C_{4h}$ to $C_{2h}$ is followed with a considerable decreasing of the HOMO orbitals energies which for the total electron spectrum constitutes

$\Delta E^{RU}$ = 226.16 kcal/mol. Taking as a whole, the data presented in Fig. 4 shows that UHF consideration of the behavior of $sp^2$-OSMs spinorbitals gives a common picture reflecting the decreasing (increasing) HOMO (LUMO) orbitals energies and degenerate orbitals splitting. The splitting value is different for orbitals of both the same molecule and different ones. The value will be discussed further in details in Section 7.2.

Coming to the second issue, we are facing the problem that for 'singlet-ground-state' fullerene $C_{60}$, (5, 5) NGr and (4, 4) SWNT the total square spin misalignment $\Delta \hat{S}^2$ is quite considerable and depends on the number of atoms in total and edge atoms with dangling bonds in the two last cases, in particular. Correctly eliminated from edge atoms, $\Delta \hat{S}^2$ can be characterized by an average value attributed to one C-C bond. Such per atom value constitutes ~0.08e for all the cases instead of zero for, say, benzene molecule which is a closed-shell molecule. In full consistence with the finding, as known, phosphorescence of benzene, both in free and condensed state, has not been observed until now while it is easily fixed for $C_{60}$ under different conditions (see review [85]) and nanosize graphene quantum dots. [86] Benzene is the most neutral solvent that resists to any photophysical event while fullerene $C_{60}$ [22, 87] and nanographenes [88] are highly active for photodynamic therapy and so forth.

Issue 3 has a direct connection to that one of molecular SOC phenomenology. The total number of effectively unpaired electrons $N_D$ is the measure of either UHF-induced radicalization of the molecule with even number of electrons or enhanced radicalization of the odd-electron radicals. It successfully plays the role of molecular chemical susceptibility [20-22] while its fraction $N_{DA}$ on atoms carries the responsibility for the atom chemical activity, atomic chemical susceptibility, distribution of which over the molecule atoms is identical to that of the atom free valence [89] (see Fig. 5 for fullerene $C_{60}$).

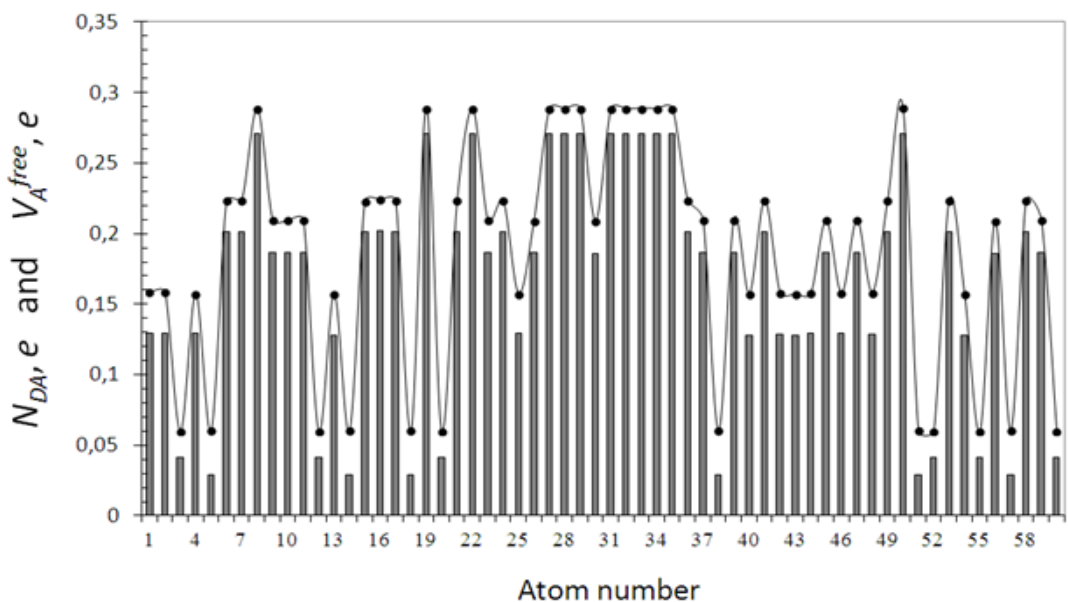

**Figure 5.** Atomic chemical susceptibility $N_{DA}$ (histograms) and free valence $V_A^{free}$ (curves with dots) over atoms of $C_{60}$ fullerene (UHF calculations). [21]

Empirical observation of para- or ferromagnetism of species with even number of electrons is usually attributed to SOC effects. [80] So called 'paradiamagnetism' of fullerene $C_{60}$ molecules was fixed at low temperature in crystalline state. [90, 91] Providing the kind permission of S.V.Demishev, Fig. 6 presents a confined view of the results obtained. As seen in panel a, the molecules, once exhibiting diamagnetic behavior up to T=40K, change the latter for paramagnetic one, 1/T, below the temperature. The observed 'paradiamagnetism' directly evidences spin mixing. Shown in panel b demonstrates that paramagnetic magnetization has a standard dependence on the magnetic field but with the Lande g-factor lying between 1 and 0.3. The strong reduction of the factor clearly indicates that electrons responsible for the observed magnetization are bound. The most amazing finding is shown in panel c, demonstrating the availability of three g-factor values revealed in the course of magneto-optical study at pulsed magnetic field up to 32 T in the frequency range $v$ = 60–90 GHz at $T$ = 1:8K. If two first features, namely, paradiamagnetism and deviation of the g-factor values from those related to free electrons, were observed in other cases as well, the third one is quite unique and is intimately connected with the electron structure of fullerene $C_{60}$.

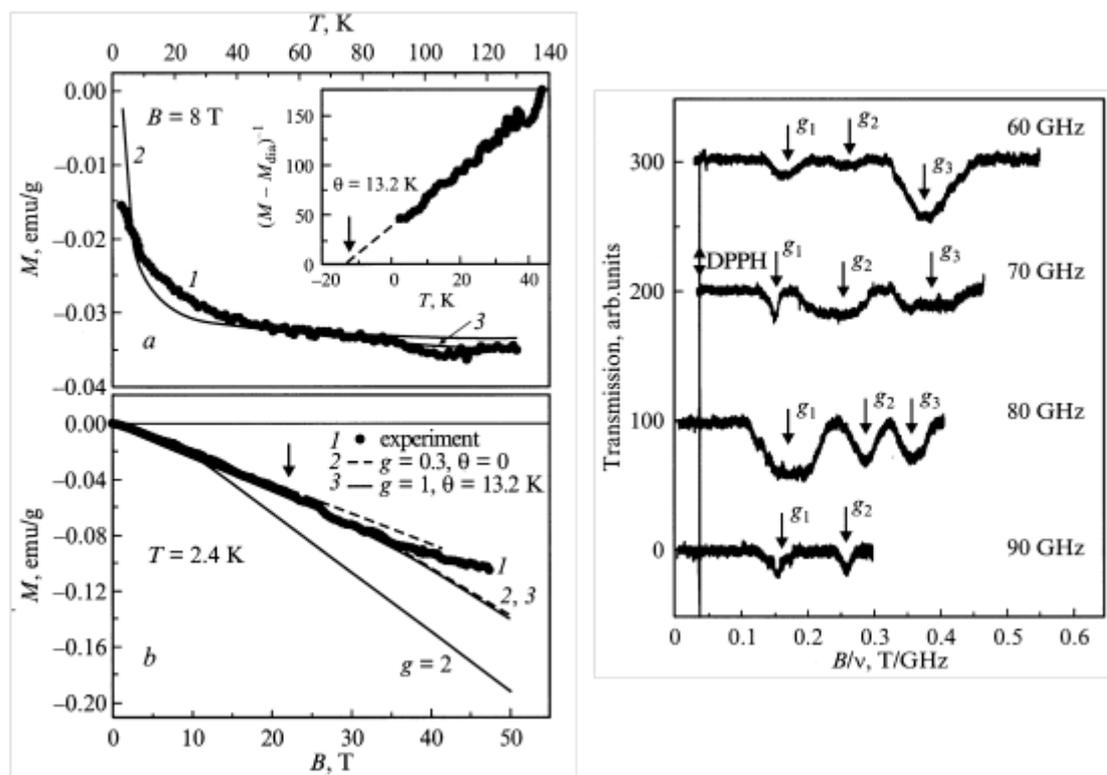

**Figure 6**. Magnetization of $C_{60}$ molecule in crystalline state. [91] Temperature (left top) and field (*left bottom*) dependence of magnetization for $C_{60}$-TMTSF-2$CS_2$ molecular complex. *1* . experimental data for $M(T)$ and $M(B)$, *2* and *3* . simulations of $M(T)$ and $M(B)$ (see details in [91]). Inset in left top shows temperature dependence of magnetization in coordinates $(M - M_{dia})-1 = f(T)$. Right. ESR absorption lines in $(ET)2C_{60}$ molecular complex at $T$ = 1:8K (by kind permission of S.V.Demishev).

Related to paramagnetic part, the feature should be attributed to a peculiar behavior of the molecule odd $p_z$ electrons while the other three valence electrons of each carbon atom are $sp^2$-configured and involved in the formation of spin-saturated σ bonds. Due to open-shell character of the molecule, part of the odd electrons, of the total number 9.87$e$, are unpaired and fragmentarily distributed over the molecule atoms (see Fig. 7). A color image of the molecule shown in Fig. 7a gives a view of this distribution. The relevant spin density on each of the 60 atoms is given by histogram. The total spin density equals zero while its negative and positive values are symmetrically distributed. As seen in the figure, the distribution clearly reveal well configured compositions of the odd electrons ('local spins' [92, 93]) with identical positive and negative spin-density values within the latter. Going over the distribution from the highest to the lowest density, one can distinguish two hexagon-packed sets, 12 singles and three sets of pairs covering 12 atoms each. All these compositions are clearly seen in the color image of the molecule. Evidently, the compositions will differently respond to the application of magnetic field thus mimicking the behavior of a molecule with different atoms. This allows for exploring the atomic expression for the evaluation of Lande g-factors related to each set of the local spin configurations of the molecule in the following form [94]

$$g = 1 + \frac{J(J+1) - L(L+1) + S(S+1)}{2J(J+1)}. \qquad (20)$$

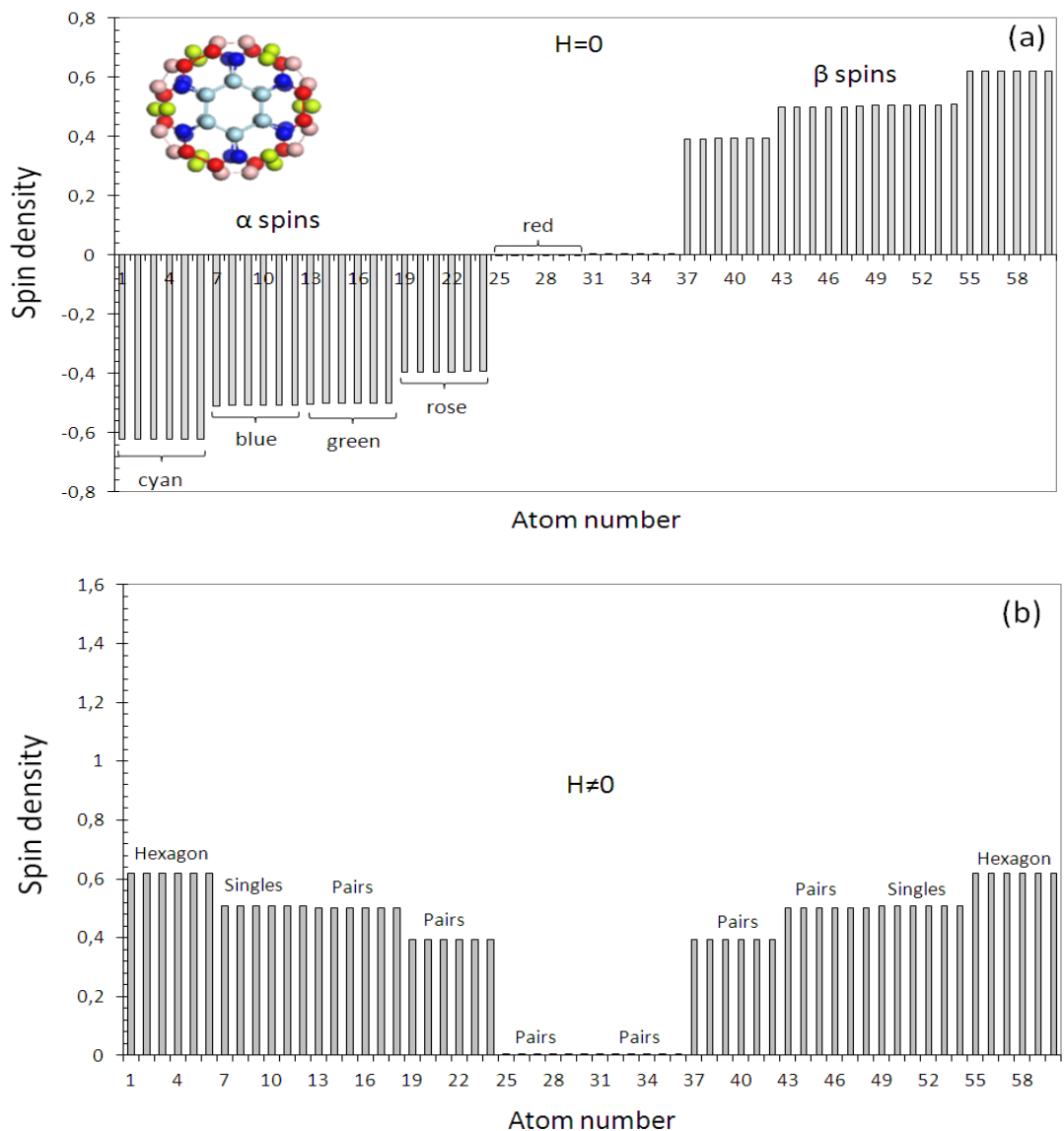

**Figure 7**. Spin density distribution over $C_{60}$ atoms in the absence (a) and presence (b) of magnetic field. Insert exhibits the color image of the local spin distribution over the molecule atoms following notations in panels a and b (UHF calculations).

. Application of magnetic field disturbs the antiferromagnetic regularity of the outlined local spins towards ferromagnetic one (see Fig. 7b) that corresponds to the maximum of magnetic ordering. If experimentally observed, the magnetic properties of fullerene molecules should reflect a variability of g-factor caused by different configuration sets of spins. The corresponding values of the factor can be evaluated by the following way. [94]

1. *Hexagon of upright local spins*.

The total spin $S = 3$, the total orbital momentum of the hexagon $p_z$ electrons $L = 6$. Consequently, the total angular momentum $J$ of the hexagon has seven components: (6+3), 8, 7, 6, 5, 4, (6-3). Usually, the lowest energy state corresponds to the minimum value of the total angular momentum that is $J =3$ in the case. Following Eq. 20, the relevant Lande factor $g_6$=0.25.

2. *Pairs of upright local spin*.

The total spin $S = 1$, the total orbital momentum of pair $p_z$ electrons is $L = 2$ and the total angular momentum $J$ has three components: (1+2), 2, (2-1). Taking the least value component, obtain the relevant Lande factor $g_2$=0.5

3. *Local spin of singles*. For a single $p_z$ electron, the total spin $= \frac{1}{2}$, the orbital angular momentum $L = 1$, while the total angular momentum $J$ has two components: 3/2 and 1/2. According to Eq. 20, the lowest energy term corresponds to $J$ = ½ so that the Lande factor $g_1$=0.66.

Thus obtained g values are collected in Table 1 alongside with experimental data. As seen in the table, experimental $g_1$, $g_2$, and $g_3$ can be attributed to singles, pairs, and hexagons of local spins, respectively. The calculated and experimental values are in a good consent. The absence of exact coincidence is evident due to a few reasons among which two next are the most important. 1) Eq. 20 is too simplified; its application allowed exhibiting quasi-polyatomic structure of the fullerene molecule related to different configurations of local spins while the exact g values determination is much more complicated problem. 2) The evaluation of g-factors was performed for ferromagnetic ordering of local spins in $C_{60}$. Experimentally, the studied crystals showed 'paradiamagnetism' exhibiting above T=40K diamagnetic behaviour that is continued as paramagnetic one below the temperature. Dominating diamagnetism is definitely provided with σ electrons that are in majority. Local spins related to unpaired $p_z$ electrons can be seen on the diamagnetic temperature-stable background only due to specific temperature dependence. Realization of the 1/T paramagnetic dependence in practice is caused by free rotation of ferromagnetically spin-configured molecules occurred at very low temperatures. [95]

Actually, when such a rotation is forbidden, as it is in the case of narrow graphene ribbons (molecules) terminated by hydrogen atoms and rigidly fixed with respect to immobile substrate, [96] the ferromagnetic behaviour of the molecules, having the same explanation as was suggested above for fullerene $C_{60}$, has been clearly observed. As shown, graphene baffles, representing a part of the original graphene sheet in direct contact with the substrate and dividing disks of hydrogenated graphene, may be of different width which occurred to be crucial for the magnetization observed. Thus, the latter is absent when the width W is zero zero (bulk graphene) and becomes well observable at W≈10 nm. When W is growing, the magnetization gradually falls down. The size dependence of the magnetization is caused by the molecular mechanism of the latter and is well explained by decreasing of exchange integrals when the size increases. [32] The magnetization is of ferromagnetic order and is connected with the behavior of local spins of graphene molecules in magnetic field. Figure 8 displays local spin behavior in the absence and under application of magnetic field relatively to the (5, 5) NGr molecule. As seen in the figure, the local spin distribution in zero field is not ordered antiferromagnetically while the total spin density is zero. This is one of the other topological shape-size effects related to condensed composition of benzenoid rings. [63,97] The field application aligns the local spins along the field thus promoting a 'ferrodiamagnetism' observation.

**Table 1**. g-Factors of fullerene $C_{60}$

| Calculated | | Experimental [91] | |
| --- | --- | --- | --- |
| Attribution | Value | Attribution | Value |
| Hexagons | 0.25 | $g_3$ | 0.19 ± 0.01 |
| Singles | 0.50 | $g_2$ | 0.27± 0.02 |
| Pairs | 0.66 | $g_1$ | 0.43 ±0.03 |

Set out in this section gives a clear vision of practically exact identity of molecular properties attributed to SOC effects in molecular physics and empirically supported UHF peculiarities of $sp^2$-OSMs.

## 7. UHF SOC parameters of $sp^2$ nanocarbons

### 7.1. A confine collection of necessary relations. Parameters' formalism

Traditionally, until now a set of standard SOC parameters has involved two values subjected to experimental verification, namely, energy splitting $\Delta E_{spl}^{SO}$ and the rate of intersystem crossing (related to singlet-triplet and vice versa transitions mainly) $k_{ISC}$ as well as theoretically introduced SOC constant $a^{SO}$. In practice all the three values are determined not by solving either Dirac-Coulomb-Breit, Eq. 15, or Dirac-Fock, Eq. 16, equations but getting first the relevant nonrelativistic problem solution after which considering $H^{SO}$ Hamiltonian mainly perturbationally, Eq. 7. [73, 77] Commonly, some effective Hamiltonian $\widehat{H}_{eff}^{SO}$ is considered.

One of the simplest and least demanding approaches is to take the two-electron contributions to the SOC into account through screening of the nuclear potential:

$$\widehat{H}_{eff}^{SO} = \frac{1}{2m^2c^2}\sum_I \sum_i \frac{Z_{i,I}^{eff}}{\hat{r}_{Iil}^3}\hat{l}_{Iil}\cdot\hat{s}_{il} \ . \tag{21}$$

In this one-electron one-center spin-orbit operator, $I$ denotes an atom and $il$ an electron occupying an orbital located at center $I$. Likewise, $\hat{l}_{Iil}$ and $\hat{s}_{Iil}$ label the angular momentum and spin of electron $il$ with respect to the orbital origin at atom $I$. The summation over electrons includes only the open shell of an atom $I$ with azimuthal quantum number $l$.

Supposing that electrons are moving in central field and substituting the fraction under the sum by the relevant static potential $V$, we get

$$\widehat{H}_{eff}^{SO} = \frac{1}{2m^2c^2}\sum_I \sum_i \frac{1}{r_{il}}\frac{\partial V}{\partial r_{il}}\hat{l}_{Iil}\cdot\hat{s}_{il} \tag{22}$$

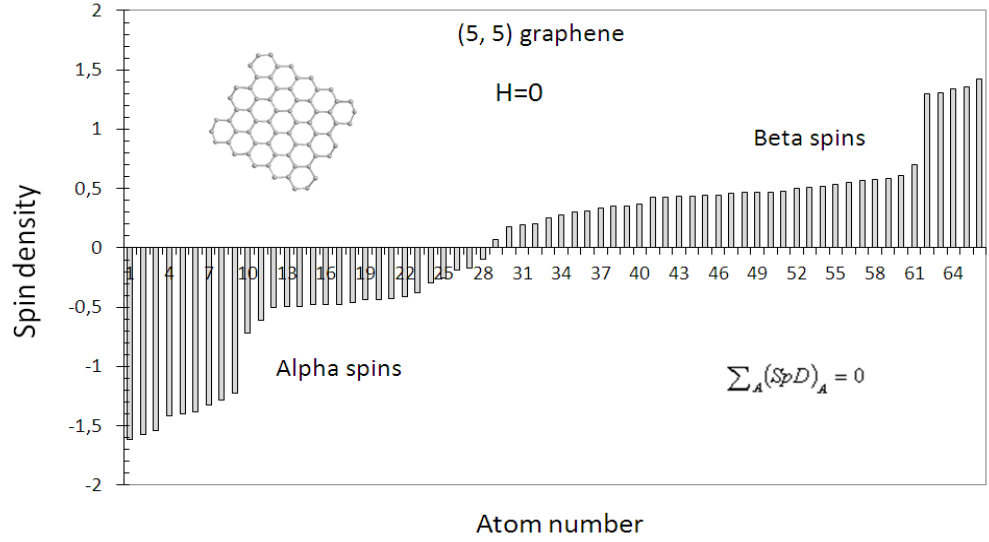

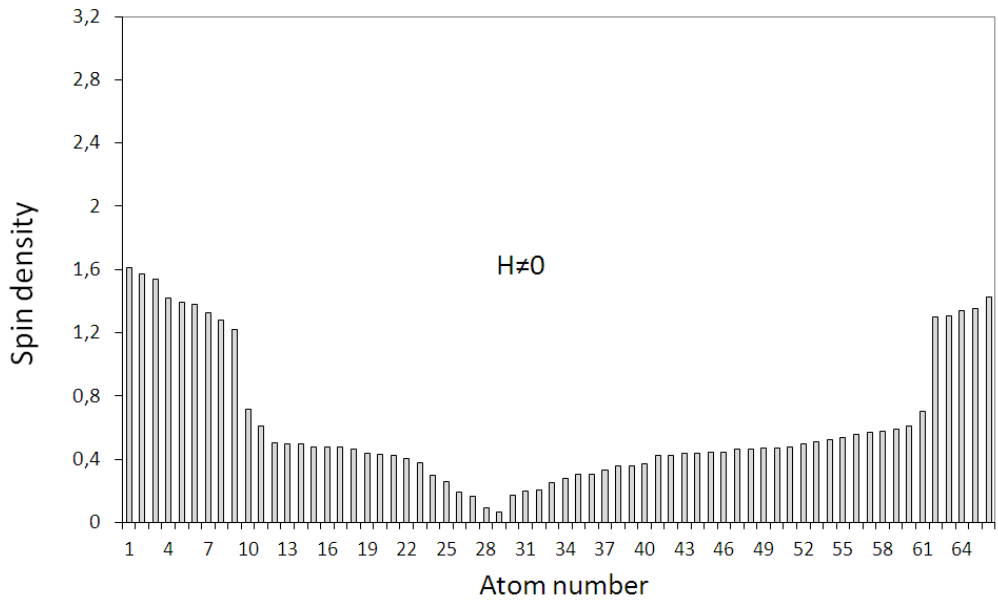

**Figure 8**. Spin density distribution over atoms of the (5, 5) NGr molecule (see insert) in the absence (a) and presence (b) of magnetic field (UHF calculations).

which can be further approximated as

$$\hat{H}_{eff}^{SO} = a^{SO}(\hat{S} \cdot \hat{L}). \quad (23)$$

In its turn, following the golden rule approximation, the ISC rate $k_{ISC}$ for the irreversible decay of an initial state $|i\rangle$, coupled by the perturbation $\sum_f H^{SO}(|i\rangle\langle f| + |f\rangle\langle i|)$ to a set of final states $|f\rangle$, is given by

$$k_{ISC} = 2\pi \sum_f |H_{if}^{SO}|^2 \delta(E_i - E_f). \quad (24)$$

Important for the following are two more approximate expressions. The first is related to eigenvalues of Eq. 7 than can be written in the form

$$\varepsilon_{SO} = a^{SO} \cdot \frac{1}{2}\{J(J+1) - L(L+1) - S(S+1)\}. \quad (25)$$

The normal $\varepsilon_{SO}$ term are related to the minimum projection of the total momentum $J$. The second presents a well known Lande interval rule

$$\varepsilon_{SO}(J) - \varepsilon_{SO}(J-1) = a^{SO}J. \tag{26}$$

Beside these standard SOC parameters, three new ones, namely, $\Delta E^{RU}$, $\Delta\langle\hat{S}\rangle^2$, and $N_D$ should be added for the set of SOC parameters to be complete. Since the meaning of the latter values and their computation have been repeatedly described in what follows we shall concentrate on the calculation of two standard SOC parameters $\Delta E^{SO}_{spl}$ and $a^{SO}$ that are the most frequently discussed with respect to SOC observation in molecules. The values evaluation will be done on the basis of output data of the UHF solutions.

### 7.2. UHF-based determination of $\Delta E^{SO}_{spl}$ and $a^{SO}$ SOC parameters

According to Eq. 23 and 26, the constant $a^{SO}$ can be determinate by two ways. In the first case, it concerns the evaluation of the potential gradient. In the second case, it just accompanies the determination of the energy splitting related to particular spinorbitals.

The first way readily addresses us to the bond dissociation and/or breaking. Since the effect of SOC on the molecule energy is remarkable but not too high, the main contribution to the energy is provided by the central Coulomb field due to which the potential gradient in Eq. 23 can be substituted by that one related to the total molecule energy. Main UHF characteristics, which accompany stretching the ethylene C-C bond up to 2Å, are shown in Fig. 9a. Equilibrium C-C distance constitutes 1.326Å and 1.415Å in singlet and triplet states of the molecule, respectively. As seen in the figure, as stretching of the bond increases, energies $E_{sg}(R)$ and $E_{tr}(R)$ approach each other up to quasidegeneracy, which is characteristic for biradicals and which is necessary for an effective SOC. [73] Simultaneously, the number of effectively unpaired electrons $N_D$, which is zero until C-C distance reaches $R_{crit}$, starts to grow manifesting a gradual radicalization of the molecule as the bond is stretched as well as exhibiting the transformation of the molecule behavior from closed-shell to open-shell one when $R_{crit}$ is overstepped.

Since we deal with the only C-C bond, the constant $a^{SO}$ can be expressed as [94]

$$a^{SO} \approx \frac{\hbar^2}{2m^2c^2}\left(\frac{1}{R}\left(\frac{d}{dR}E_{SO}(R)\right)\right) \tag{27}$$

where $R$ corresponds to the current C-C bond length. The expression in the inner brackets describes the force acting on the bond under stretching. The deviation of the force under progressive stretching the ethylene C-C bond is presented in Fig. 9b. At the beginning it proceeds linearly starting, however, to slow down when $R$ is approaching $R_{crit}$. A clearly seen kink is vivid in the region. For comparison, the curve with horizontal bars presents the force caused by stretching a single C-C bond of ethane, $R_{crit}$ for which constitutes 2.11Å, [43] that is why the molecule remains closed-shell one within the interval of C-C distances presented in the figure. As seen, the force is saturated at the level of 110 kcal/(mol*Å) that is close to the kink position on the ethylene curve. It is quite reasonable to suggest that the force excession over this value in the latter case is caused by the closed-open shell transformation of the ethylene molecule over 1.4Å due to which the force excess is caused by the SOC. Consequently, this excess can be considered as $dE_{SO}(R)/dR$ that is presented by the gray-ball curve in the figure.

Using the curve values in the C-C distance interval from 1.40Å to 1.47Å, which is typical for the C-C bond dispersion in fullerenes, CNTs, and graphene, and substituting them into Eq. 27, one can obtain the $a^{SO}$ constant laying in the interval from 15 meV to 110 meV, which is typically expected for molecules of light elements [73] and which was predicted [98] and experimentally determined [99] for graphene.

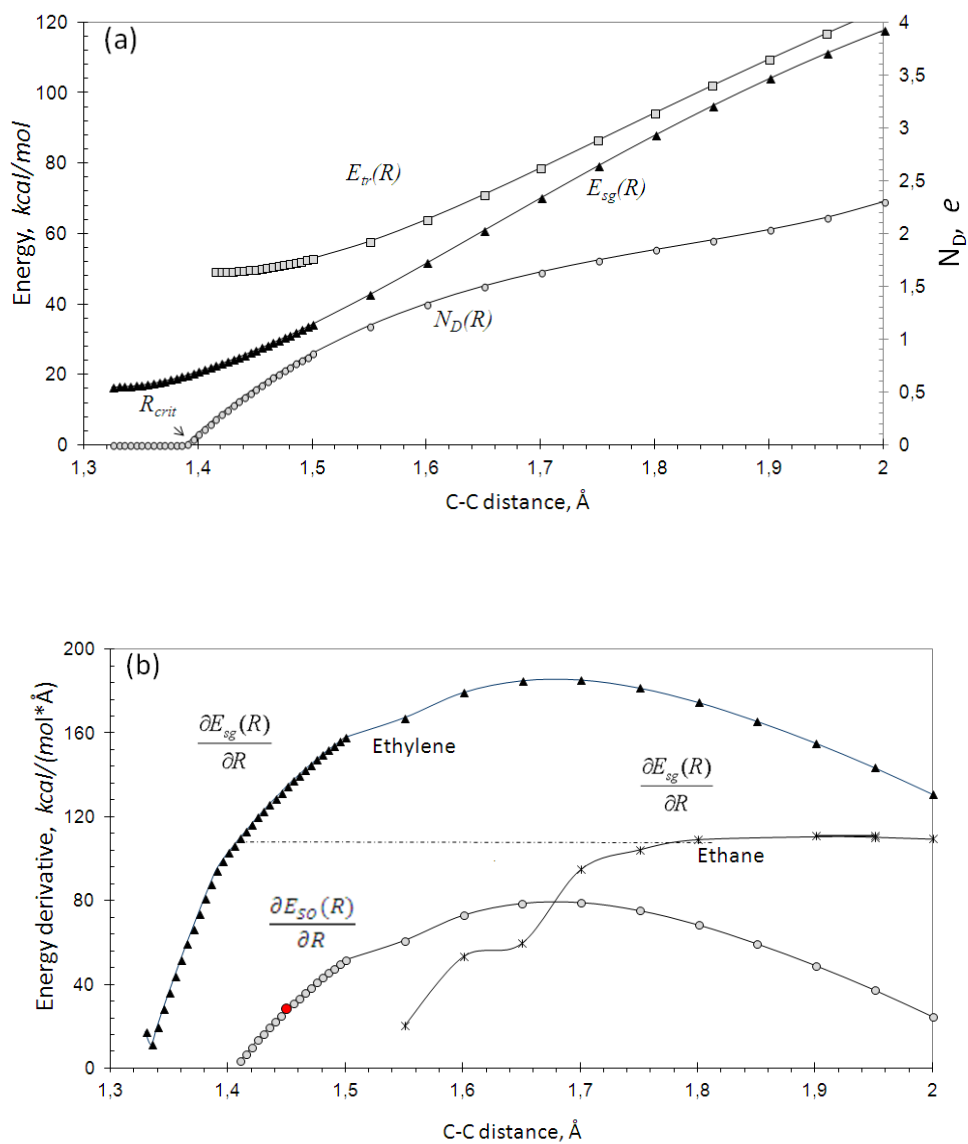

**Figure 9.** a. Energy of singlet and triplet states as well as the total number of effectively unpaired electrons $N_D$ of ethylene vs C-C distance. b. Energy derivatives vs C-C distance (see text) (UHF calculations).

It is necessary to point out as well that the $dE_{SO}(R)/dR$ force maximum amplitude for ethylene molecule is well consistent with those determined under uniaxial deformation of benzene molecule (Fig. 10) and graphene (Fig. 11) [62] due to which the outlined $a^{SO}$ constant values can be considered as typical for the whole family of $sp^2$ nanocarbons.

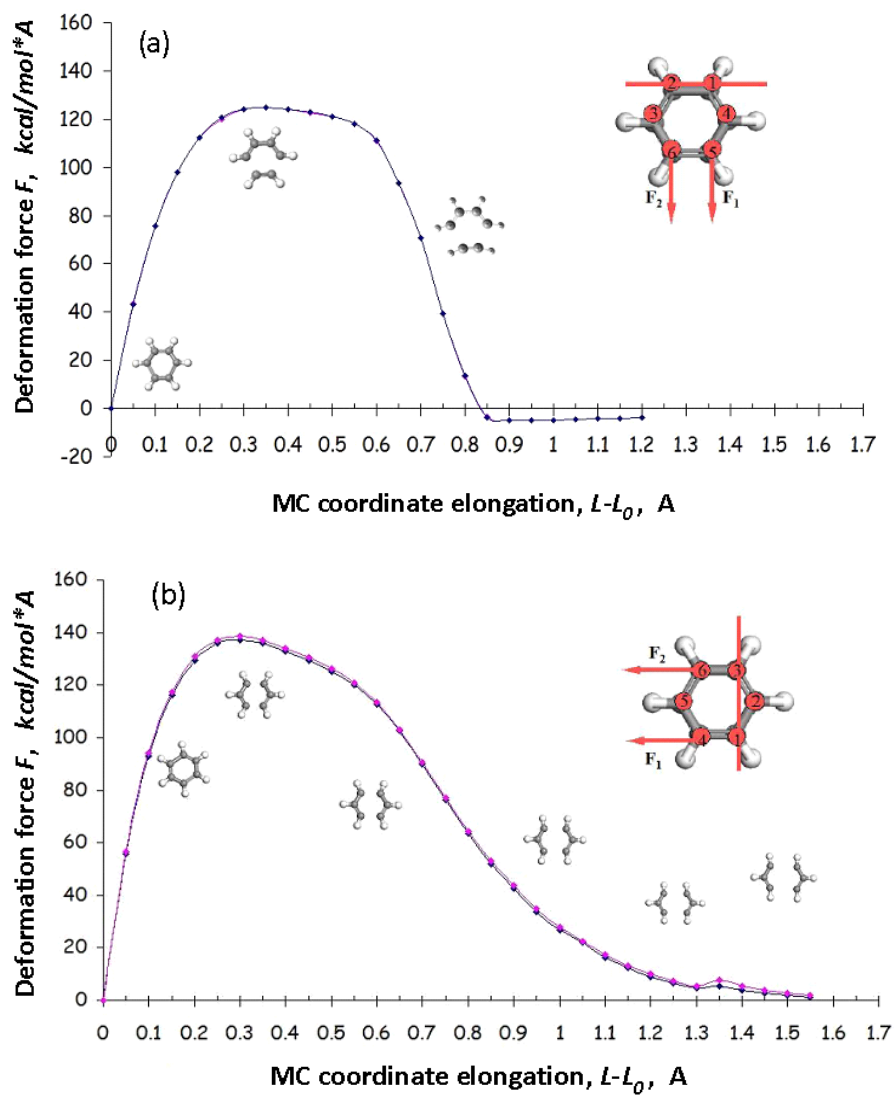

**Figure 10.** Deformation forces applied to two mechano-chemical coordinates vs the coordinate elongation at the benzene molecule uniaxial tension related to the armchair (a) and zigzag (b) modes. Inserts present the modes configuration (UHF calculation).

Presented in Fig. 10 is related to uniaxial tension of the benzene molecule. The deformation is considered in terms of mechano-chemical reaction whose description is peformed in terms of mechano-chemical coordinates (MCCs), inserted in the pool of the molecule internal coordinates by a specific way.[100] Configurations of the relevant coordinates for benzene and (5, 5) NGr molecules are shown in inserts in Figs 10 and 11. For benzene itself as well as for all the bodies constructed of condensed benzenoid units, the deformation is highly anisotropic, multimode, depending on the MCC configuration. Giving in inserts, MCCs correspond to the deformation normal to C-C bonds (panels a, armchair mode) and along the bonds (panels b, zigzag mode).[62]

As seen in Fig. 10, deformation along both MCCs of benzene is practically identical due to which applied forces F1 and F2 cannot be distinguished. In the case of (5, 5) NGr molecule in Fig. 11, six applied forces are quite different and the difference depends on the deformation mode configuration. At the same time, the force maximum values are well similar pointing on the governing role of a benzenoid unit in this mechano-chemical reaction. A detailed description of the deformation of benzene and (5, 5) NGr molecule is given everywhere.[62] Here we are interested in the deformation force applied only.

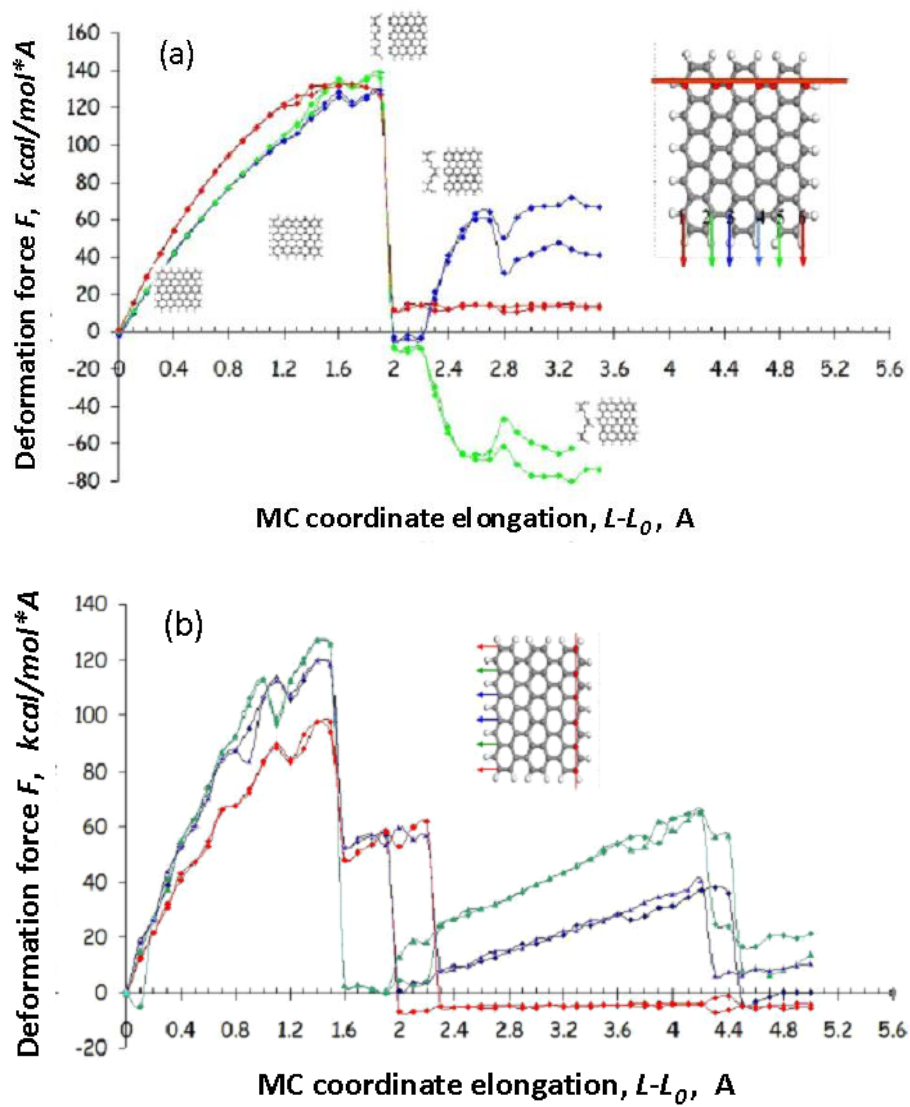

**Figure 11.** Deformation forces applied to six mechano-chemical coordinates vs the coordinate elongation at the (5, 5) NGr molecule uniaxial tension related to the armchair (a) and zigzag (b) modes. Inserts present the modes configuration (UHF calculation).

Since the equilibrium C-C bond length of benzene just coincides with $R_{crit}$ while exceeds the value in the (5, 5) NGr molecule, the forces in Figs, 10, 11 correspond to the $dE_{SO}(R)/dR$ region due to which their amplitudes should be compared with the latter one for ethylene in Fig. 9. As seen in Figs. 9-11, the force amplitude changes from 80 kcal/mol in ethylene to ~130-140 kcal/mol for both benzene and (5, 5) NGr. Indicating strengthening of the double C-C bond when going from ethylene to benzene and condensed benzenoid structures, the difference in the applied force amplitude shows as well that the previously estimated values of the constant should be considerably increased.

Following the second way, the SOC constant can be evaluated on the basis of spinorbital splittings presented in Fig. 4 by using the Lande interval rule, Eq. 26. The constant values related to the selected sets of data are presented in Table 2. Zero as the constant minimum corresponds to the absence of some spinorbitals splitting. The constant maximum corresponds to the biggest splitting related to spinorbitals presented in Fig. 4. Looking at the data, one can conclude that the constants determined on the basis of dissociation of ethylene molecule fit the interval quite well.

**Table 2**. Spin-orbit coupling cocnstants $a_{SO}$ (UHF calculation)

| $sp^2$ nanocarbons | $a_{SO}$ constant, meV |
|---|---|
| Fullerene C$_{60}$ | 0 - 60 |
| (5, 5) NGr molecule | 0 - 230 |
| (4, 4) SCNT (fragment) | 0 - 60 |

### 7.1. About the state-of-art of the SOC insight into the $sp^2$ nanocarbons

Following from the said in the previous sections, polyradicalism and the resulting corollaris as UHF evidence of SOC, on the one hand, and energy splitting, SOC constant and intersystem crossing rate, on the other, are two sets covering UHF- and standard SOC realities. Combining these, at first glance, so different phenomena will certainly form in the future the basis of the molecular theory of OSMs. However, so far they have been considered separately without tracing its inner connection. Let us have a brief look what do these two sets look like with respect to $sp^2$ nanocarbons.

     Polyradicalism of fullerenes has not been so far of a long history and have been still considered by the majority of computationists as nothing to have in common with closed-shell fullerene molecules. Nevertheless, the author activity towards open-shell character of the molecules started in 2003 (see [22] and references therein) was supported first by Stück et al in 2011 [24] and then Scuseria et al just recently. [25] A profound analysis of the open-shell nature of fullerenes performed in [25] as well as the demonstration of both UHF and GHF approaches
brought the authors very close to make the final conclusion about spin-orbit nature of the peculiarities observed. However, this last step has not been done. New insight on fullerenes as SO OSMs undoubtedly will open new pages in their fascinating history. So far the only paper [99] has been found dealing standardly with the SOC in fullerenes comparatively with that in CNTs and graphene at the basis of perturbation theory $H^{SO}$ approach.

     View on CNTs as polyradicals is not popular among the large scientific community as well, which prefers to attribute them to closed-shell systems. This causes a quite small set of the relevant publications. [22, 23, 26-29] In contrast, a great attention is given to the standard SOC study in CNTs. [80, 101-112] The studies, mainly computational, are concentrated on the fine structure of the electronic bands of translationally symmetric CNTs depending on their transverse size (curvature), chirality and other structural features. There is an obvious impact of analogous studies of graphene attributing CNTs to curved graphene. Experimentally, the SOC constant $a^{SO}$ value constitutes 0.37±0.2 meV [103] and 0.15 meV. [108] Computationally evaluated the constant values are range from 0.076 meV to 2.4 meV. Those are dependent on the tube diameter and chiral structure.

     In contrast to fullerenes and CNTs, both polyradicalism [23, 30-42] and standard SOC study of graphene [98,99,101, 113-129] are quite rich. The main pulse of interest is connected with the Rashba-Sheka effect [130] adapted to 2D crystals (see a comprehensive discussion in [131]). The studies were concentrated on SOC-stimulated changes in the graphene crystal band structure, particularly, on the conductive bands splitting in some points of the Brillouin zone. The studied splittings are ranged from ~100 meV to ~10 meV. [99] Combining with molecular magnetism of graphene, [96] the available data evidence a significant role of SOC in electronics and magnetronics of graphene molecules and crystals.

1. Conclusion

The paper presents a new concept of the peculiarities of open-shell molecules attributing them to the effect of spin orbit coupling. The main issues involve two sets of characteristics. The first covers spin contamination of the molecule ground state, $\Delta\langle S\rangle^2$, depraving the latter from the exact spin multiplicity, and effectively unpaired electrons of $N_D$ total number. The characteristics are proven to be directly associated with the molecule behavior in reality. The second set involves standard SOC parameters such as splitting energy $\Delta E_{spl}^{SO}$ , rate of intersystem crossing $k_{ISC}$ , and SOC constant $a^{SO}$. The integration of the two sets in a single one, seemingly paradoxical due to different origin of their members, is proven by the adequacy of the UHF results, forming the first set, to the same ones obtained in the framework of the relativistic molecular theory, on the one hand, as well as the consistency of the second set values calculated by using UHF resultant data with those obtained by standard SOC approach, on the other. The UHF formalism, formally nonrelativistic, produces spin-associated data provided with nondiagonal Coulomb elements of the relevant $f^\alpha$ and $f^\beta$ Fock matrices, related to α and β spins. Comparing the Unrestricted Hartree-Fock and Dirac-Fock equation structure it may be suggested that the identity of their results is enclosed in two-electron-integral implementation of the computing algorithm. The estimation of the accuracy of such a presentation of the Dirac-Fock equation as well as of the relativicity inherent in the UHF formalism is still waiting for its clarification. In spite of the remaining spots, the SOC origin of the open-shell molecules peculiarities can be considered as well substantiated. Evidently, this new viewpoint opens a large way for the reconsideration of expectations concerning such important $sp^2$-OSMs as fullerenes, CNTs, and graphene. Evidently strongly affecting their chemistry, the SOC contribution into $p_z$ electron behavior will significantly influence their physics as well.

**Acknoledgements**
The author greatly appreciate fruitful and stimulating discussions with V. Sheka, J. Karwowski, I. Mayer, E. Orlenko, P. D'yachkov, S. I. Vinitski, A. Gusev, Yu. P. Rybakov, E. Brandas, M. Nucimento, and D. Mukherjee as well as L. Gross, L. Buchinsky, and S.V. Demishev,  for kind permition to use experimental data at our discretion and L. Kh. Shaymardanova and N.A. Popova for participating in the calculations related to the deformation of benzene and (5, 5)NGr molecules. The work was performed under financial support of the Peoples' Friendship University of Russia, grant: 022203-0-000.